\documentclass[twocolumn]{article}

\usepackage{lmodern}
\usepackage{authblk}
\usepackage{amsmath}
\usepackage[colorlinks=true, allcolors=cyan]{hyperref}
\usepackage{graphicx,xcolor}
\usepackage{cite}
\usepackage[top=2.5cm]{geometry}

\title{\LARGE \bf Optical parametric multi-pass cell amplifier}
\author[1,*]{Supriya Rajhans}
\author[1]{Nikolas Rupp}
\author[1]{Esmerando Escoto}
\author[1]{Arthur Sch\"{o}nberg}
\author[2]{Dominic Laumer}
\author[2]{Malte Sumfleth}
\author[3]{Issam Abdallah}
\author[3]{Bastian Manschwetus}
\author[4]{Caroline Juliano}
\author[1]{Nikan Javid}
\author[4]{Cord L. Arnold}
\author[2]{Tais Gorkhover}
\author[2]{Markus Drescher}
\author[3]{Robert Riedel}
\author[1]{Ingmar Hartl}
\author[1,5,6]{Christoph M. Heyl}
\author[1]{Tino Lang}

\affil[1]{Deutsches Elektronen-Synchrotron DESY, Notkestraße 85, 22607 Hamburg, Germany}
\affil[2]{Universität Hamburg, Institut für Experimentalphysik, Luruper Chaussee 149, 22761 Hamburg, Germany}
\affil[3]{Class 5 Photonics, Notkestraße 85, 22607 Hamburg, Germany}
\affil[4]{Lund University, 22100 Lund, Sweden}
\affil[5]{Helmholtz Institute Jena, Fröbelstieg 3, 07743 Jena, Germany}
\affil[6]{GSI Helmholtzzentrum für Schwerionenforschung GmbH, Planckstraße 1, 64291 Darmstadt, Germany}

\affil[*]{Corresponding author: supriya.rajhans@desy.de}

\begin{document}
\twocolumn[
  \begin{@twocolumnfalse}
    \maketitle

\begin{abstract}
Ultrafast lasers with simultaneously high average and peak power have become indispensable for driving a multitude of applications, including high-harmonic generation, strong-field physics, and particle source applications. Both parametric amplifiers and post-compressed Ytterbium-lasers have emerged as prime platforms to meet these demands.
While multi-pass cell (MPC) based post-compression offers broadband output with high beam quality, it provides limited wavelength tunability and suffers from temporal contrast degradation. Conversely, optical parametric amplifiers (OPAs) provide spectral tunability and high temporal contrast but they are limited by low pump-to-signal conversion efficiency and spatial beam inhomogeneities. Here, we introduce the Optical Parametric Multi-Pass Cell Amplifier (OPMPC), a hybrid architecture that overcomes the limitations of both schemes. Our approach utilizes two non-collinearly intersecting MPCs providing broadband parametric amplification of the seed pulses and complete idler removal after each pass through the crystal, thereby suppressing back-conversion.
We experimentally demonstrate a record pump-to-signal power conversion efficiency of 43\% using a 1030\,nm pump at a 1\,kHz repetition rate with a pulse energy of 174\,µJ. The amplified signal at 1500\,nm exhibits excellent beam quality, power and spectral stability and is compressed to 48\,fs, demonstrating a new platform for ultrafast pulse generation.
\end{abstract}
\vspace{5 mm}
\end{@twocolumnfalse}
  ]
\maketitle
\section{Introduction}

The development of ultrafast laser sources that combine high average power, high pulse energy, and shortest pulse duration is a key enabler for a wide range of applications in science and technology. High-harmonic generation (HHG), strong-field physics, and laser-driven plasma acceleration all benefit from pushing the performance boundaries of these systems. Although Ti:Sapphire lasers have long served as the standard for high-energy, few-cycle pulses, their operation at high energy levels is typically limited to low repetition rates ($<$ 1 kHz) \cite{eichner_cryogenic:25,Moulton:86,Wolter:17}. In recent years, significant progress has been made in the power and energy scaling of Ytterbium (Yb)-based amplifier systems, which now routinely provide up to kilowatts of average power and tens of millijoules of pulse energy at high repetition rates spanning from the kilohertz (kHz) to the megahertz (MHz) range \cite{Dietz:20,Pergament:25,mueller_18-kw_2018,russbueldt:10}. However, these systems typically emit pulses with duration of a few hundred femtoseconds to about one picosecond, centered at a wavelength of 1030 nm. To unlock their full potential for ultrafast applications, highly efficient conversion of these high-power pump pulses into both ultrashort duration and different spectral regions is a critical ongoing challenge. 
A powerful toolkit of established techniques has emerged in order to address these demands. Post-compression techniques are well established for converting the high average power of picosecond laser systems into ultrashort pulses with durations ranging from a few optical cycles to several tens of femtoseconds. Among these, post-compression using multi-pass cells (MPC) offers distinct advantages, including high throughput, the preservation of excellent beam quality, and spatio-spectral homogeneity due to their quasi-waveguiding properties \cite{Hanna:21,viotti_multi-pass_2022,Kaumanns:21,Pfaff:23,Rajhans:23}. For wavelength conversion, optical parametric amplifiers (OPAs) provide a versatile platform. They enable generation of tunable high power ultrashort laser pulses across the visible, near- and mid-infrared (NIR/MIR) spectral regions and support high pulse energies \cite{Byer:93,Cerullo:03,Fattahi:14,Kretschmar:20}. 
These two technological pillars, post-compression and optical parametric amplification, have each established their own domain of excellence in pushing the limits of ultrafast laser performance. The direct combination of the strengths of these approaches represents a compelling and logical frontier. Recently, a major step in this direction was reported with the introduction of an optical parametric process within a multi-pass cell concept in two simultaneous, independent studies\cite{Rajhans_cleo:25,Thannheimer_cleo:25}, followed by a comprehensive experimental validation of this concept \cite{nagele:25}. This work demonstrated the fundamental feasibility of merging these two technologies, showcasing an architecture that leverages the MPC's ability to suppress the idler beam, circumventing the traditional bandwidth and efficiency limits of conventional OPAs. 
However, the full potential of this hybrid concept, in particular the path to higher pulse energies, shorter pulses and even higher conversion efficiencies close to the theoretical quantum limit as well as the systematic exploitation of the MPC's quasi-waveguiding properties to ensure spatio-spectral homogeneity remains an open and critical challenge.

Here, we address these challenges and report the experimental demonstration of an Optical Parametric Multi-Pass Cell Amplifier (OPMPC), designed to unlock the full potential of combining MPC and OPA technologies. Our results are grounded in a deep theoretical understanding of the OPMPC concept, developed through extensive 3+1D simulations that include second- and third-order nonlinearities together with spatio-temporal effects within the MPC and the birefringent nonlinear crystals. 

Our model reveals the path towards near-quantum-limited efficiency and confirms the inherent scalability of the architecture. We validate our approach experimentally by generating 75\,µJ, 48-fs pulses centered at 1500\,nm with a record 43\,\% pump-to-signal power conversion efficiency. The incorporation of a novel non-collinear OPMPC geometry in the experimental setup enables direct idler suppression and flexible pass-by-pass group delay compensation between pump and signal. We demonstrate excellent spatio-spectral homogeneity, nearly diffraction-limited beam quality (\textit{M²} $<$ 1.3), and outstanding power stability (0.2\,\% rms). These results demonstrate the successful generation of high-energy, tunable pulses with record pump-to-signal conversion efficiency and excellent beam quality, establishing OPMPC as a new paradigm for next-generation ultrafast sources.

    \begin{figure*}[!tbp]
\centering
\includegraphics[width=0.75\linewidth]{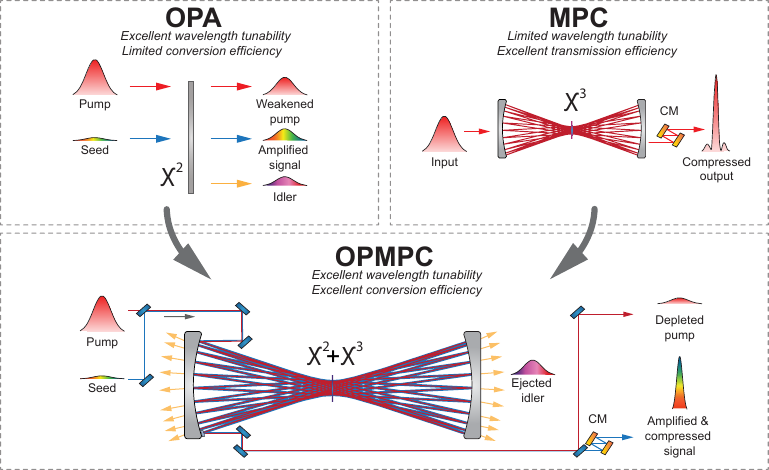}
\caption{Conceptualization of the OPMPC scheme, based on the combination of optical parametric amplification (OPA) and multi-pass cell (MPC) approaches, combining the respective advantages of both schemes. In the OPMPC, pump and seed beam co-propagate within the MPC passing a nonlinear crystal placed at the focus. The generated idler is ejected after each crystal pass e.g. via using a suitable mirror coating. After amplification, the signal beam is separated and compressed utilizing chirped mirrors (CM). Here, $\chi^{2}$ and $\chi^{3}$ represent the second-order and third-order nonlinear susceptibilities, respectively.}
\label{fig:concept_figure}
\end{figure*}

\section{Concept}

Both MPC-based post-compressed laser systems as well as OPAs are constituting foundational technologies for ultrafast science, each approach with its own advantages and limitations. As discussed in the previous section, MPC-based post-compression has allowed the efficient generation of few-cycle pulses from high power picosecond lasers, resulting in an output having both high peak and average powers. However, large compression ratios lead to a degradation in the temporal contrast of the pulse \cite{Escoto:22}. This originates from inherent spectral modulations in SPM-based spectral broadening and higher-order dispersion, resulting in a significant portion of the energy in pre-/post-pulses. 

Apart from degradation of the temporal contrast, another limitation of the multi-pass cell is modest spectral tunability despite attempts to address this challenge \cite{Balla:23,Cichelli:25}.
In contrast, wavelength flexibility and broad spectral coverage are inherent advantages of OPA technology, determined by the phase-matching conditions and the transparency window of the nonlinear crystal. However, OPAs face a significant and well-documented limitation in their amplification efficiency. The pump-to-signal power conversion efficiency for state-of-the-art femtosecond OPAs is typically capped at around 20\,\%, while the efficiency normalized to the quantum limit generally remain below 45\,\% (see Fig.~\ref{fig:overview_figure_opa}). Furthermore, achieving the ultra-broad bandwidths needed for few-cycle, high-peak-power pulses requires birefringent crystals in non-collinear geometries. In such configurations, spatial and temporal walk-off effects become unavoidable, fundamentally compromising the spatio-spectral homogeneity of the amplified signal \cite{Lang2013,Eichner:22}.
 
These limitations, while seemingly restrictive, reveal a complementary nature. The primary weakness of MPCs is the core strength of OPAs and vice versa. This complementary relationship suggests the opportunity to synthesize a new architecture that compensates for the weaknesses of each technique while combining their individual strengths. Figure~\ref{fig:concept_figure} illustrates the fundamental principle of the OPMPC. In the depicted simplified collinear representation, a second-order nonlinear crystal is placed at the focus of a Herriott-type multi-pass cell. Both pump and signal beams propagate collinearly through the OPMPC, amplifying the signal pulses with each pass through the nonlinear crystal. 
A major advantage of this approach is that the idler beam, generated in the parametric process, can be easily filtered out each pass, e.g. by using suitable mirror coatings which transmit the idler. If the length of the nonlinear crystal is kept shorter than the effective coherence length of the nonlinear process in a single crystal pass, the overall back conversion into the pump pulses is suppressed due to the absence of the idler at the beginning of each crystal pass. This key mechanism, combined with an optimized pump-signal temporal overlap for each pass and the mitigation of spatial hole burning through the MPCs quasi-waveguiding property, enables high amplification efficiencies. Furthermore, despite the spatial walk-off effects in the birefringent crystal, these same quasi-waveguiding properties suppress spatio-spectral inhomogeneities, ensuring a high-quality output beam. Finally, the nonlinear and instantaneous nature of the parametric amplification process ensures a high temporal contrast. Amplification is confined strictly to the temporal overlap window of the pump and signal pulses, preventing pre- or post-pulse amplification. This results in an output pulse contrast far superior to  that of a pulse generated using SPM-based post-compression, relaxing the need for additional temporal cleaning techniques. 

To guide our experimental implementation and explore the properties of the OPMPC concept, we developed a comprehensive 3+1D numerical model using the Chi3D software framework \cite{lang_chi23d_2022}. This model utilizes a split-step Fourier algorithm and incorporates all relevant physical processes: all relevant second-order nonlinear interactions, material dispersion, absorption, spatial and temporal walk-off, and third-order Kerr effects (self-phase modulation and self-focusing).
The simulations are carried out for a collinear OPMPC geometry and the parameters are chosen to closely replicate our experimental constraints. A single potassium titanyle arsenate (KTA) crystal, cut for type II (o→eo) phase-matching for a signal wavelength of 1500\,nm (41.4° in the XZ-plane), is placed at the common focus of collinearly propagating signal and idler. Self-phase modulation and self-focusing are included, taking into account a nonlinear refractive index n$_2$ of $1.7 \times 10^{-19}$\,{\textrm{m}$^2/$W} \cite{jansonas_interferometric_2022}.\par

For the simulations, the input seed spectrum is derived directly from our experimentally measured white-light continuum (see Fig.~\ref{fig:conversion_eff_amplified_spectrum}(b)), with a pulse energy of about 10\,nJ. Its spectral phase is approximated by considering the linear dispersion, calculated from the Sellmeier coefficients of the materials in the beam path (white light YAG crystal and 7.5\,mm fused silica from the mode-matching telescope), with an additional group-delay dispersion (GDD) of 390\,fs² to account for the nonlinear phase accumulated during the SPM-driven white-light generation process, resulting in an estimated second-order dispersion of 265\,fs\(^2\) and third-order dispersion of 1500\,fs\(^3\). The input pump spectrum is derived from the measured spectrum with a spectral phase of 5700\,fs\(^2\) applied in order to match the frequency-resolved optical gating (FROG) retrieved pulse duration of $\sim$200\,fs. 

We consider Gaussian spatial beam profiles for seed and pump pulse with a waist radius of 0.66\,mm and 0.54\,mm at the nonlinear crystal, respectively. After each crystal interaction, the beams complete one half-round-trip within the MPC, propagating 566\,mm from the MPC center at which the nonlinear crystal is placed to an MPC mirror with a 2\,m radius of curvature and back to the crystal for the next pass. This defines an MPC configuration with corresponding parameters $N=14$, $k=5$ \cite{viotti_multi-pass_2022}. The model rigorously incorporates all linear losses. A base loss of 0.75\,\% is applied per pass to account for the crystal's anti-reflection coatings. The total simulated losses are consistent with the linear transmission experimentally measured for the pump pulse after 20 passes. Specifically for the signal beam, which is more sensitive to absorption in the spectral region of interest, the model includes atmospheric attenuation. This is modeled using the HITRAN database for water vapor absorption at 50\,\% relative humidity, which includes the corresponding phase dispersion as calculated by the Kramers-Kronig relations \cite{gebhardt_impact_2015}. Finally, for these initial simulations, a standard literature value for the effective nonlinear coefficient, d$_\textrm{eff}$, of 2\,pm/V for KTA is used with a thickness of 0.8\,mm.

Our simulations confirm the core advantages of the OPMPC architecture. The model predicts a maximum pump-to-signal conversion efficiency of 51.4\,\% approaching the theoretical quantum limit of 68.7\,\% for this process, taking into account the losses for each pass. The calculated per-pass evolution of conversion efficiency and pump depletion under this optimal condition is shown in  Fig\,\ref{fig:conversion_eff_amplified_spectrum}(a). Furthermore, the simulation shows excellent beam quality and spectral homogeneity. The overall gain bandwidth appears limited only by the phase-matching bandwidth of a single crystal pass, rather than being narrowed by the cumulative second-order nonlinear interaction over all 20 passes. In addition, the model shows  SPM-driven spectral broadening of the pump and signal pulses within the nonlinear crystal, analogous to traditional post-pulse compression MPCs, which further reduces the transform-limited pulse duration of the amplified signal to sub-30\,fs. In the following section, these simulation results are directly compared with our experimental findings.

\section{Results}

\begin{figure}[!htpb]
\centering
\includegraphics[width=\linewidth]{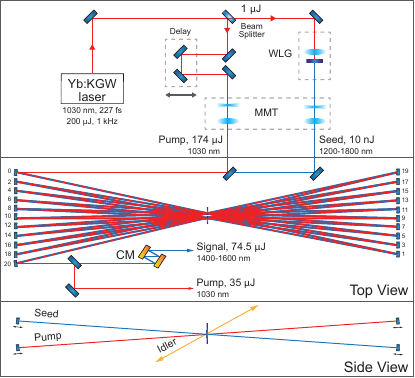}
\caption{Schematic of the experimental setup with the top and side view of the OPMPC scheme (WLG: White light generation, MMT: Mode-matching telescope, CM: Chirped mirrors).}
\label{fig:schematic_experiment}
\end{figure}
To experimentally validate the concept, we implement an OPMPC scheme analogous to that described in the previous chapter. The initial experimental layout, however, differs from the simulated optimum configuration as presented in Fig.\ref{fig:concept_figure}. The selected non-collinear geometry provides several practical advantages, which are discussed in detail in this section. 

The experimental setup is depicted in Fig.~\ref{fig:schematic_experiment}. The pump laser (Pharos PH1-20) provides 227\,fs pulses centered at 1030\,nm wavelength with a pulse energy of 200\,µJ at a repetition rate of 1\,kHz. A small fraction of the input pulse energy is picked off to generate the OPMPC seed by filamentation in a YAG crystal, producing a stable white-light continuum.
\begin{figure*}[!htb]
\centering
\includegraphics[width=0.98\linewidth]{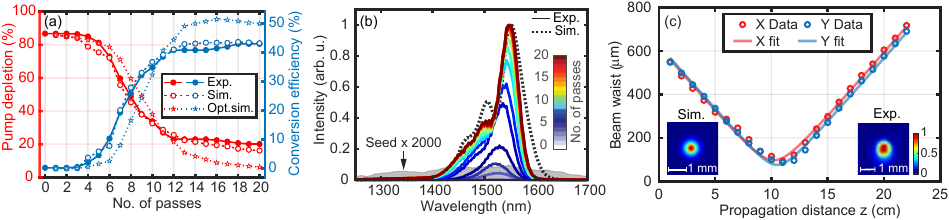}
\caption{(a) Simulated (Sim.) and experimentally measured (Exp.) depletion of the pump, and conversion efficiency from the pump to the signal beam as a function of the number of passes in the OPMPC. We also display the optimized simulations (Opt.sim.) results for the maximum conversion efficiency reached with the theoretical  d$_\textrm{eff}$, of 2\,pm/V for KTA. 
(b) Experimentally measured amplified spectrum after each pass through the OPMPC with the colorbar indicating the number of passes. The gray shaded area shows the input seed spectrum magnified for visualization alongside the simulated amplified signal spectrum corresponding to the simulated (Sim.) conversion efficiency plot in (a).  (c) Beam quality measurement of the signal output yielding an \(M^2\) value of $1.29\times1.01$. The simulated and measured signal beam profile are shown as insets.}
\label{fig:conversion_eff_amplified_spectrum}
\end{figure*}
The maximum pump pulse energy available for the OPMPC is 174\,µJ. A delay stage in the pump arm allows fine adjustment of the relative timing between pump and seed before the OPMPC. Both beams are then sent through telescopes to match their spatial modes to the MPC, using cell parameters similar to those considered in the simulation.\par

In contrast to the simulation, we employ a non-collinear OPMPC geometry in which the pump and seed propagate in two independent MPCs and are spatially and temporally overlapped in the nonlinear crystal. This separation enables the use of dedicated off-the-shelf dielectric mirrors optimized for pump and seed beams. Although a collinear setup in a conventional Herriott configuration is more compact and generally less alignment-sensitive, it imposes feasible but demanding requirements on the dielectric mirror coatings: the reflectivity and spectral group delay must be precisely engineered to simultaneously support pump and seed, mitigate temporal walk-off, and suppress the idler. In contrast, the non-collinear OPMPC layout enables flexible temporal alignment between pump and seed by introducing small translation stages for each pass. Moreover, in this geometry, idler suppression occurs naturally: the idler exits the cavity due to the non-collinear beam geometry. As shown in the schematic (Fig.~\ref{fig:schematic_experiment}), the non-collinear OPMPC has two linear rows of 10 mirrors on each side with the seed and pump mirrors on top and bottom, respectively. The placement of the seed and pump mirror rows are reversed on the opposite side to support the geometry. By using a separate mirror for each beam reflection inside the MPC, a cell alignment supporting two foci at the center of the cell is achieved. One KTA crystal is placed at each focus for parametric amplification in each pass. The thickness of the KTA crystal is 1.5\,mm with anti-reflective coating for both the pump and seed wavelengths.\par 

Each pass is aligned to optimize the spatial and temporal overlap between the pump and the seed. After 20 passes through the KTA crystals, we achieve a record pump-to-signal power conversion efficiency reaching 42.9\,\% (pump into the MPC, signal out after 20 passes)  as shown in Fig.~\ref{fig:conversion_eff_amplified_spectrum}(a). The system is clearly reaching saturation for this particular configuration. Figure ~\ref{fig:conversion_eff_amplified_spectrum}(a) also shows the pump depletion with amplification in each pass. The power meter head is positioned at the output of the OPMPC setup for this measurement, thus the total linear losses resulting from the 20 passes in the OPMPC are included in all the measurement points. In order to acquire data corresponding to a specific pass, the signal beam is blocked during subsequent passes. The signal output power measured after 20 passes is 74.5\,mW corresponding to a pulse energy of 74.5\,µJ. The pump is depleted by 80\% in the 20 passes taking into account the cell losses which amount to 13.3\%. The amplified spectrum measured over the 20 passes is shown in Fig.~\ref{fig:conversion_eff_amplified_spectrum}(b).  In order to check the spatial quality of the output beam, an \(M^2\) measurement is carried out resulting in $1.29\times1.01$ in the x- and y-axes respectively,  as shown in Fig.~\ref{fig:conversion_eff_amplified_spectrum}(c). The x- and y-axes corresponds to the horizontal and vertical axis of the OPMPC setup. The \(M^2\) fit is carried out considering \(1/e^2\) beam waists. The beam profile of the signal output is also shown as an inset in this plot. The quasi-waveguiding property of the MPC results in spatially homogeneous profiles, as discussed in the concept section. 
To achieve a quantitative match with the measured conversion efficiency, the d$_\textrm{eff}$ in the simulation was adjusted to 1.4\,pm/V considering the experimentally used 1.5\,mm KTA crystal. Following this adjustment, the simulation accurately reproduces all key experimental findings, including the pump-to-signal conversion efficiency [Fig:\ref{fig:conversion_eff_amplified_spectrum} (a)], spectral shape [Fig:\ref{fig:conversion_eff_amplified_spectrum} (b)], and beam quality [Fig: \ref{fig:conversion_eff_amplified_spectrum} (c)]. 

\begin{figure}[!b]
\centering
\includegraphics[width=\linewidth]{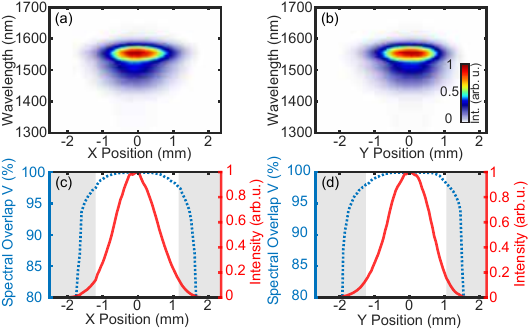}
\caption{(a) Spectrally resolved beam profile measurement along the (a) x-axis and (b) y-axis, and the corresponding spectral overlap parameter $V$ in (c) and (d), respectively. The areas where the intensity is lower than \(1/e^2\) of the maximum are shaded in gray. } 
\label{fig:spatio-spec_data}
\end{figure}

In order to verify the spatial beam properties in greater detail, a spatio-spectral homogeneity measurement is carried out. Figure~\ref{fig:spatio-spec_data}(a) and (b) show the spectra measurements along the x- and y-axes respectively, along with the corresponding spectral overlap parameter $V$ as defined in Ref. \cite{Weitenberg:17}.The output displays excellent spatio-spectral homogeneity with the spectral overlap $>$ 97\,\% in both dimensions.

\begin{figure}[!b]
\centering
\includegraphics[width=\linewidth]{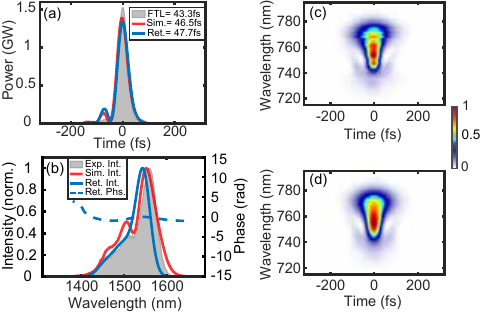}
\caption{(a) Simulated and measured compressed temporal output pulse profile retrieved from the FROG traces along with the FTL pulse. (b) Simulated spectrum, retrieved spectrum and spectral phase along with the measured spectrum. Measured (c) and retrieved (d) FROG traces of the compressed signal output. } 
\label{fig:compression_data}
\end{figure}

The amplified spectrum supports a Fourier-transform-limited (FTL) duration of 43\,fs. To verify the compressibility of this pulse, a wedge reflection of the output signal is sent onto chirped mirrors providing a negative group delay dispersion of -170\,fs$^2$. The recorded and retrieved FROG traces are shown in Fig.~\ref{fig:compression_data} (c) and (d). The retrieval error is 0.26\% for a grid size of $4096\times1498$. The retrieved temporal pulse profile is shown in Fig.~\ref{fig:compression_data}(a) with a full-width at half-maximum (FWHM) of 48\,fs, close to the FTL duration. The estimated peak power, taking into account the signal output pulse energy of 75\,µJ is 1.3\,GW. The simulations carried out with the d$_\textrm{eff}$ of 1.4\,pm/V also results in a pulse duration close to the experimental measurements as shown in Fig.~\ref{fig:compression_data}(a). The corresponding simulated spectral shape is plotted along with the measured spectrum, retrieved spectrum, and spectral phase from the FROG measurements in Fig.~\ref{fig:compression_data}(b). The compression is expected to improve further by optimizing the third order dispersion in the compressor.

\begin{figure}[!t]
\centering
\includegraphics[width=0.95\linewidth]{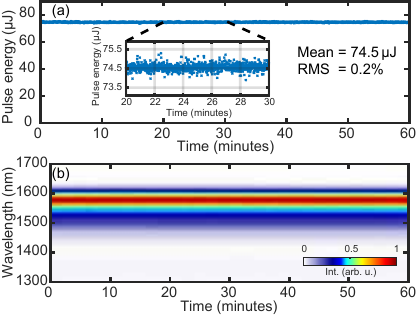}
\caption{(a) Output power measurement as a function of time measured with a thermal power head recorded after 20 passes. (b) Corresponding output spectrum measured as a function of time with an integration time of 1\,ms.}
\label{fig:power_spectral_stability}
\end{figure}

The OPMPC system is very stable in terms of output power, pulse energy, and spectral width. A measurement of this stability over several minutes is shown in Fig.~\ref{fig:power_spectral_stability}(a) with the output signal power measured with a thermal sensor head having a response time of 0.6\,seconds. The measurements are sampled at 0.7\,seconds and the corresponding pulse energy plotted as a function of time gives an rms stability of 0.2\,\%. 
Fig.~\ref{fig:power_spectral_stability} (b) displays the spectral stability using an integration time of 1\,ms.

 \par  

\section{Discussion}

This work demonstrates an ultrafast laser source that combines two widely used approaches supporting high average and peak powers: OPAs and MPC-based post-compression. The objective is to reduce the limitations inherent to each technique while preserving their advantages. Our comprehensive 3+1D simulations confirm that this objective is possible. High pump-to-signal conversion efficiencies approaching the quantum limit can be achieved, with the remaining constraints arising primarily from optical losses within the system. The OPMPC scheme supports broadband spectra with good spatial homogeneity as shown in  Fig.~\ref{fig:conversion_eff_amplified_spectrum} and Fig.~\ref{fig:compression_data}. The pulses can be well compressed with appropriate dispersion compensation. 

\begin{figure}[!b]
\centering
\includegraphics[width=0.9\linewidth]{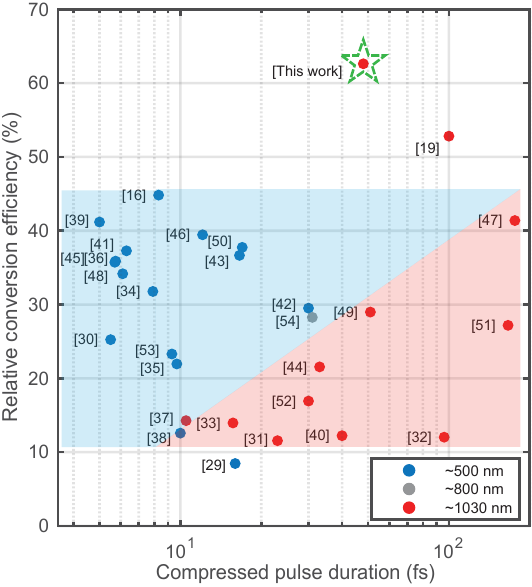}
\caption{Pump-to-signal power conversion efficiency normalized to the quantum limit as a function of the compressed output pulse duration as reported by different experimental optical parametric amplification works. The data covers different input pump wavelengths as described in the legend. The result discussed here is highlighted by a green star. } 
\label{fig:overview_figure_opa}
\end{figure} 
The promising simulation results are successfully validated experimentally, as reported in the previous section. The setup delivers 43\,\% pump-to-signal conversion efficiency, including cell losses, alongside high beam quality, stability, and reliable pulse compressibility. The output pulses are compressed to 48\,fs, close to the Fourier-transform limit of the broadened spectrum. 

Our initial implementation also indicates several routes for improvement. Enhanced anti-reflective coatings could substantially improve overall system efficiency, while the implementation of a tighter focusing geometry would likely enable more effective pump depletion. Employing thinner nonlinear crystals offers a pathway to even broader usable bandwidths. Enclosing the system to maintain low humidity would reduce absorption losses arising in air, and more precise pump-beam characterization on each pass would support finer optimization of pump–signal overlap.\par

Although there is clear room for future optimization, the reported performance  represents state-of-the-art results as shown in Fig.~\ref{fig:overview_figure_opa}. This overview figure presents the pump-to-signal conversion efficiency normalized to the quantum limit (relative conversion efficiency) as a function of the compressed pulse duration \cite{Killi:06,Adachi:08,Moses:09,Chalus:09,Gu:09,Herrmann:09,Emons:10,Schultze:10,Deng:12,Fattahi:12,Rothhardt:12,Hong:14,Matyschok:14,Hoppner:15,Puppin:15,Shamir:15,Prinz:15,Batysta:16,Mundry:17,Harth_2018,Mero:18,Mecseki:19,Windeler:19,Feng:20,Hrisafov:20,Kretschmar:20,Xu:20,nagele:25}, and covers the parameter space achieved by various OPA studies. Most OPAs pumped near 500\,nm (blue shaded region) exhibit relative power conversion efficiencies below 45\,\%. In contrast, the majority of all systems pumped near 1030\,nm achieve relative efficiencies below 30\,\%, with only a single report approaching 40\,\%. The only two reported systems capable of exceeding a 50\,\% relative power conversion efficiency are based on a combined OPA and MPC approach, the work reported in Ref. \cite{nagele:25} and the here discussed result. Our experimental results demonstrate a markedly superior relative efficiency ($>$60\,\%), in agreement with the 3+1D simulations, setting a new record for ultrafast parametric sources. Our approach thus provides a compelling platform for the development of a new generation of ultrashort, high-average power, high-peak power, and wavelength tunable laser sources.

\section{Conclusion and Outlook}
The Optical Parametric Multi-pass Cell Amplifier (OPMPC), which strategically combines the advantages of multi-pass cells and optical parametric amplification, is conceptually presented and experimentally demonstrated. Using a 1030-nm pump, the OPMPC scheme enabled us to reach a record pump-to-signal conversion efficiency of 42.9\,\% at a 1500\,nm signal wavelength, while maintaining good spatial beam quality and spatio-spectral uniformity. The output pulse is compressible to 48\,fs, significantly shorter than the 227\,fs pump, with notable stability in both spectrum and power. The immediate steps to improve the performance of the present system are clearly outlined in the previous section. Beyond these near-term optimizations, the proposed scheme provides more generally a versatile framework for advanced ultrafast laser source development. 
Extending the concept into the mid-infrared e.g. by amplifying the idler rather than the signal represents a particularly promising avenue.
Short-pulse sources in these spectral regions are essential for industry-relevant technologies, including atmospheric light detection and ranging (LIDAR), trace-gas sensing, and prospective free-space optical communication \cite{hlavatsch:22}. Furthermore, these sources can play a key role in scientific applications such as nonlinear phononics for inducing electronic and magnetic phase transitions in complex materials \cite{Nicoletti:16}. A key limitation in these fields remains the limited average power of existing mid-infrared sources. Scaling the architecture presented here offers a clear pathway towards overcoming this bottleneck and realizing high-performance systems in the near future.

In addition, energy and average-power scaling of the OPMPC architecture should be feasible in close analogy to MPC systems, with principal constraints arising from the damage threshold of the nonlinear medium and the cell optics. Numerical simulations already indicate that OPMPC configurations can support pulse energies at the millijoule level \cite{Rajhans_cleo:25}. Achieving such scaling would render these systems highly attractive for demanding applications such as high-harmonic generation and laser-driven plasma acceleration.

Last but not least, the adaptation of the OPMPC concept to other nonlinear processes offers exciting opportunities for nonlinear optics and ultrafast laser science. Promising candidates include harmonic generation, difference frequency generation and four-wave mixing.
\par


\subsubsection*{Funding and Acknowledgments} We acknowledge DESY (Hamburg, Germany), a member of the Helmholtz Association HGF, for support and the provision of experimental facilities. We further acknowledge support from the Deutsche Forschungsgemeinschaft (DFG, German Research Foundation) – project 545612524 as well as from the German BMFTR, project 13N16678.

\subsubsection*{Disclosures} The authors declare no conflicts of interest.

\subsubsection*{Data availability}Data underlying the results presented in this paper are not publicly available but may be obtained from the authors upon reasonable request.
Except where otherwise noted, all figures and images in this document are the original work of the authors and are licensed under a \href{https://creativecommons.org}{CC-BY 4.0 License}.



\bibliography{mybibliography}


\end{document}